\documentclass{aa}
\def\msol{{M}_{\odot}}

\def\lesssim{\,\lower 1mm \hbox{\ueber{\sim}{<}}\,}
\def\grsim{\,\lower 1mm \hbox{\ueber{\sim}{>}}\,}
\def\ueber#1#2{{\setbox0=\hbox{$#1$}%
  \setbox1=\hbox to\wd0{\hss$ #2$\hss}%
  \offinterlineskip
  \vbox{\box1\box0}}{}}
\input psfig
\begin{document}

   \thesaurus{03     % A&A Section 3: extragalactic
%              ( 11.03.4; %{\bf Galaxies: clusters: individual:} $\ldots$
                11.09.3; %intergalactic medium
                12.03.3; %Cosmology: observations
                12.04.1; %dark matter
%                11.01.2; %Galaxies: active
                13.25.2)} % X-rays: galaxies
%               13.25.3)} % X-rays: general

\title{Distant clusters of galaxies: X-ray properties and their relations} 

   %%  \subtitle{Test}

   \author{	Sabine Schindler 
	}

%   \offprints{Sabine Schindler}
   \institute{
               Astrophysics Research Institute, 
               Liverpool John Moores University, 
	       Twelve Quays House,
               Egerton Wharf,
               Birkenhead CH41 1LD,
               U.K.;
               e-mail: {\tt sas@staru1.livjm.ac.uk}
              }
   \date{}
\titlerunning {Distant clusters of galaxies}
   \maketitle

   \begin{abstract}
We investigate the evolution of clusters of galaxies in a sample of distant
clusters with redshifts between 0.3 and 1.0. We show 
the abilities and limitations of combined ROSAT and ASCA data 
to draw cosmological conclusions.
For the first time bolometric luminosities, 
masses, gas masses, gas mass fractions, and iron masses are derived in
such a distant sample in a consistent way. 
We compare these quantities with the
corresponding quantities in nearby samples. Furthermore, we analyse
relations between these quantities and the 
gas temperature, metallicity and the morphological parameters and
compare them with relations in nearby samples. Fits to all 
relations with power law functions are given. 
We find relations between X-ray luminosity, temperature, mass, gas
mass, core radius and $\beta$, similar to those found
in nearby clusters. Furthermore, we find gas mass fractions
increasing with radius, with the effect being stronger in less massive
clusters. Within errors we
find no evidence for evolution in any of the quantities nor 
in any of the relations. These results favour a low $\Omega$
universe, but without strong constraints. 
We point out how promising the next generation of X-ray
satellites XMM, CHANDRA and ASTRO-E are for cosmological studies with
clusters. From the new observations 
primarily two measurements are required: detection of
more distant clusters and measurements of the spectral and spatial
parameters with much higher precision.

   \end{abstract}

%
%________________________________________________________________
%

\section{Introduction}

Distant clusters of galaxies provide important diagnostics for
cosmology. They are the largest bound systems in the universe and
as such can be used to test theories of the origin and the evolution of
structure. Different cosmological models predict different relations
between quantities obtainable from X-ray observations like the
bolometric X-ray luminosity, the intra-cluster gas temperature, the
total mass and the gas mass (e.g. Cavaliere et al. 1998b). 
A different
evolution of these relations is predicted as well by different
theories (see e.g. Bower 1997; Voit \& Donahue 1998). 
The evolution of the cluster mass and the related
quantity temperature
depend e.g. sensitively on the cosmological parameters $\Omega$ and
$\Lambda$ (see e.g. Oukbir \& Blanchard 1992; Cen \& Ostriker
1994). Even the amount of dispersion in relations can reflect
different cluster formation epochs (Scharf \& Mushotzky 1997).
Deviations from theoretical scaling relations (e.g. Kaiser 1986)
can give useful insights into physical processes going on in clusters.
 
The differences in observables predicted by cosmological theories are
becoming larger and larger with increasing redshift. Tests of theories
are therefore most powerful, when clusters with the largest possible
redshifts are used.

In this paper we analyse relations between the X-ray quantities of a
sample of distant clusters ($z=0.3-1.0$) and
the evolution of these quantities by comparing them with the
corresponding nearby results. 
Most relations are well measured for nearby clusters (e.g. the $L_X -
T$ relation: Arnaud \&
Evrard 1999; Allen \& Fabian 1998; Markevitch 1998, Reiprich 1998). 
The reasons why
only few attempts have been made to determine X-ray relations in
medium distant 
and distant clusters (e.g. Mushotzky \& Scharf 1997), despite their
importance to constrain cosmological models, are the large
uncertainties in the measurements and the paucity of clusters found at 
high redshifts. 

The sample used in this paper consists of all clusters at $z>0.5$
with measured ASCA temperature and measured morphological parameters
from ROSAT/HRI observations, complemented with 5 clusters fulfilling
the same criteria within the redshift range $0.3<z<0.5$. 
No scaling relations are used, but for the first time in such a
distant sample all quantities are derived directly from
the measured parameters: temperature, countrate or luminosity in the
ROSAT band, metallicity and morphological parameters. 
Masses and bolometric luminosities are calculated in a consistent
way and masses are determined within equivalent volumes.
Unlike our analysis, which compares comprehensively all derivable
quantities, previous studies of cluster samples up to a redshift of 1
showed only few relations of directly measured quantities.

We demonstrate what the
abilities and limitations of cosmological studies with distant
clusters are when using 
a combination of the best X-ray data available today,
spatially resolved data from the ROSAT/HRI (Tr\"umper 1983) 
and X-ray spectra from ASCA (Tanaka et al. 1994). We show that
with these data one sees similar relations as for nearby clusters, but
of course with much less accuracy. For solid cosmological 
conclusions we have to wait for the new missions XMM, CHANDRA and
ASTRO-E with their much enhanced  
collecting area, spatial and spectral resolution.

Other tests for cosmological models would be the investigation of 
the evolution of the
luminosity function or the temperature function (see e.g. Burke et
al. 1997; Henry 1997) or using the distribution of clusters to
determine the correlation function (e.g. Mo et al. 1996; 
Guzzo et al. 1999). 
For these tests the selection function of the sample must be
known very well. As the selection function for the sample in this
paper is completely unknown, we make no attempt to determine any of these
functions.

The paper is organised as follows. After a description of the data
sources and analysis methods (Sect. 2)  we show various relations in
Sect. 3: relations of total mass, gas mass and iron mass (Sect. 3.1),
relations with temperature (Sect. 3.2) and relations with the X-ray
luminosity (Sect 3.3). Finally, Sect. 4 gives our summary and conclusions.
Throughout this paper we use $\rm{H}_0 = 50$ km/s/Mpc and $\rm{q}_0=0.5$.

\section{Data}

We select the most distant clusters with published ASCA temperatures
which were also observed with the ROSAT/HRI (see Table~\ref{tab:input}). 
The sample includes all the clusters
meeting this criterion above a redshift of $z=0.5$. These are 6
clusters. For the redshift
range $z=0.3-0.5$ only a few clusters of the many observed ones were
selected. Including more of these low redshift clusters would have
resulted in a sample heavily dominated by clusters at redshifts $z=0.3-0.5$
which would have made it useless for a study of the evolution of clusters.   
As all the clusters in the sample 
above a redshift of $z=0.5$ show the gravitational
lensing effect we chose 5 clusters in the redshift range $z=0.3-0.5$,
which show a gravitational lensing signal as well as meet the above
mentioned criteria. In total the sample consists of 11 clusters.

Obviously, the sample is not complete in any sense, therefore no
analyses of distribution functions can be made with it. But the
analyses of correlations between the different quantities should not be
affected by the incompleteness because the clusters were detected by only 
ONE criterion, e.g. high X-ray luminosity or high mass in case of 
the lensing cluster AXJ2019+112. All other quantities, apart from this
one that was used to detect the cluster, are therefore
not biased in any way. Therefore the unknown selection function can
hardly influence the results on correlations. 

\begin{table*}[htbp]
\begin{center}
\begin{tabular}{|c|c|c|c|c|c|c|c|c|c|c|}
\hline
cluster & z & $L_{X,bol}$ & m & T & $S_0$ & $r_c$ & $\beta$ & source& $r_{tot}$&ref.  \\
        &   & ($10^{45}$erg/s)&(solar)&(keV)&& (kpc)&&counts&(Mpc) &\cr
        &      &         &                &          &&&&&& \cr
\hline
AC118       & 0.31 & 6.6 &$0.23\pm0.09$& $9.3^{+4.2}$ &1.5 &$370\pm40$ &$0.63\pm0.4$&5300&2.1& a,b,c\cr
Cl0500-24   & 0.32 & 0.6 &$0.0-1.5$  & $7.2^{+3.8}_{-1.8}$  &0.8&$30^{+470}_{-30}$&$0.4^{+0.6}_{-0.1}$&440&0.9& d,e \cr
Cl0939+4713 & 0.41 & 1.6 &$0.22^{+0.24}_{-0.22}$&$7.6^{+2.8}_{-1.6}$ &0.79&$66^{+90}_{-47}$&$0.36^{+0.9}_{-0.7}$&1100&1.0& f \cr
RXJ1347-1145& 0.45 & 21. &$0.33\pm0.10$ &$9.3^{+1.1}_{-1.0}$ &61.&$57\pm12$&$0.57\pm0.04$&2200&1.7& g \cr
3C295       & 0.46 & 2.6 & -            &$7.1^{+2.2}_{-1.3}$ &33.&$26\pm16$&$0.52\pm0.07$&680&0.8& b,h \cr
Cl0016+16   & 0.55 & 5.2 &$0.07^{+0.11}_{-0.07}$&$7.6^{+0.7}_{-0.6}$&1.6 &$283^{+59}_{-48}$&$0.68^{+0.10}_{-0.07}$&3700&1.5&i,j \cr
MS0451-0305 & 0.55 & 7.0 &$0.15^{+0.11}_{-0.12}$&$10.4\pm1.2$&2.2&$256^{+69}_{-53}$&$0.68^{+0.13}_{-0.09}$&1400&1.6& k \cr 
Cl2236-04   & 0.55 & 1.5 &$0.0^{+0.38}_{-0.0}$&$6.2^{+2.6}_{-1.7}$ &3.9&$66^{+41}_{-27}$&$0.53^{+0.18}_{-0.09}$&480&0.6& l \cr
RXJ1716+6708& 0.81 & 1.4&$0.43^{+0.25}_{-0.21}$&$5.7^{+1.3}_{-0.6}$&0.85&$56\pm50$&$0.42^{+0.14}_{-0.09}$&790&1.0&m \cr 
MS1054-0321 & 0.83 & 7.1 &0.0-0.22&$12.3^{+3.1}_{-2.2}$&0.59&$\approx500$&$0.7-1.0$&1100&$\grsim$0.5& n \cr
AXJ2019+112 & 1.01 & 1.1 &$1.7^{+1.25}_{-0.74}$&$8.6^{+4.2}_{-3.0}$  &0.33&$\approx150$&$\approx0.9$&76&0.5& o \cr

\hline
\end{tabular}
\end{center}
\caption{X-ray quantities as measured from ROSAT/HRI and ASCA
observations. The clusters (column 1) are
ordered according to redshift (column 2). Column (3), (4) and (5) list the
bolometric X-ray luminosity, the metallicity in solar units and the
temperature, respectively. In columns (6), (7) and (8) the fit
parameters of the $\beta$ model are shown, central surface brightness, 
core radius $r_c$ and the slope 
$\beta$. $S_0$ is in units of $10^{-2}$
ROSAT/HRI counts/s/$\sq\arcmin$. Columns (8) and (9) give the total
number of source counts in the ROSAT observation and the radius out to
which the X-ray emission could be traced. Column (10) denotes the references:
(a) Mushotzky \& Loewenstein 1997,
(b) Mushotzky \& Scharf 1997,
(c) Neumann \& Schindler 1999,
(d) Schindler \& Wambsganss 1997,
(e) Ota et al. 1998,
(f) Schindler et al. 1998,
(g) Schindler et al. 1997,
(h) Neumann 1999,
(i) Neumann \& B\"ohringer 1997,
(j) Hughes \& Birkinshaw 1998,
(k) Donahue 1996,
(l) Hattori et al. 1998,
(m) Gioia et al. 1999,
(n) Donahue et al. 1998,
(o) Hattori et al. 1997}
\label{tab:input}
\end{table*}

In literature different 
methods are used to determine the bolometric
luminosity and we found considerable variations from author to
author. In our analysis we
derive the bolometric luminosity in a uniform way for all the
clusters with the ROSAT data analysis software EXSAS. We start 
from the ROSAT/HRI countrate. Only for the clusters AC118 and
Cl0016+16, for which also ROSAT/PSPC observations are available, we
use the countrates from the ROSAT/PSPC observation as 
this is the more sensitive instrument and provides therefore a more
accurate measurement of the luminosity. We use the temperature,
metallicities and hydrogen column densities given in the publications
listed in Table~\ref{tab:input}. The derived bolometric luminosities
are listed there as well.

As the mass of a
cluster is increasing with radius 
masses can only be compared when derived within equivalent
volumes. The masses in literature are usually determined at arbitrary
radii and are therefore not directly comparable.
We determine all the masses at a radius $r_{500}$ which
encompasses a density 500 $\times$ the critical density 
$\rho_c(z) = 3 H_0^2 / (8\pi G) (1+z)^3$ (see Table~\ref{tab:mass}). A
comparison with the X-ray extent of the clusters (see
Table~\ref{tab:input}) shows that in most clusters the X-ray
emission could be traced out to $r_{500}$.
For the determination of the
radial dependence of the gas mass fraction we determine additionally
the mass at the radius $0.5\times r_{500}$.

To deproject the two-dimensional X-ray emission to
three-dimensional densities and masses we use the 
$\beta$ model (Cavaliere \& Fusco-Femiano 1976; Jones \&
Forman 1984)
$$S(r) = 
S_0 \left( 1 + {\left({r \over r_c}\right)}^2\right)^{-3\beta + 1/2},
   \eqno(1)
$$
where $S(r)$ is
the surface brightness at distance r, 
$S_0$ is the central surface brightness, $r_c$ is the core
radius, and $\beta$ is the slope. We use only $\beta$-models fitted to
ROSAT/HRI data for all clusters. 
A comparison with $\beta$-model parameters of
ROSAT/PSPC data shows that the ROSAT/PSPC parameters are systematically larger
for these high redshift clusters, even when they are deconvolved with
the point spread function. While for nearby
clusters the ROSAT/PSPC data give a good estimate of the mass, for clusters
at high redshifts the point spread function distorts the profile,
i.e. it flattens the profile in the centre. The
result is a $\beta$-model fit with a larger core radius in the
ROSAT/PSPC data and this is compensated by a larger $\beta$.
This larger $\beta$ yields a steeper slope in the outer parts
and thus results in a larger mass estimate at the radius considered here. 
This effect can overestimate the
mass for high redshift clusters up to almost 50\% (e.g. in
Cl0016+16 using fit parameters by Neumann \& B\"ohringer (1997) and
Hughes \& Birkinshaw (1998) or in  Cl0939+4713 using the parameters of
Schindler \&
Wambsganss (1996) and Schindler et al. (1998)). The difference depends
of course on the radius where the mass is determined.
Compared to this the effect of the point spread function of the
ROSAT/HRI on the mass is small (\lesssim 10\%).

\begin{table*}[htbp]
\begin{center}
\begin{tabular}{|c|c|c|c|c|c|c|c|c|}
\hline
cluster & z & $r_{500}$ & $M_{tot,500}$
& $M_{gas,500}$ & $f_{gas,500}$ &$M_{tot,500/2} $ & $M_{gas,500/2}$ & $f_{gas,500/2}$
\cr
&&(Mpc)&($10^{14}\msol$) &($10^{14}\msol$)& &($10^{14}\msol$)&($10^{14}\msol$) & \cr
        &      &         &                &          &&&& \cr
\hline
AC118        & 0.31&$1.35^{+0.25}_{-0.10}$&$8.1^{+3.7}_{-1.1}$&1.75&$0.22^{+0.04}_{-0.07}$&$3.3^{+1.5}_{-0.5}$&0.54&$0.16^{+0.03}_{-0.05}$\cr
Cl0500-24    & 0.32&$0.97^{+0.23}_{-0.13}$&$3.0^{+1.6}_{-0.7}$&0.25&$0.08^{+0.03}_{-0.03}$&$1.5^{+0.8}_{-0.4}$&0.07&$0.05^{+0.02}_{-0.02}$\cr
Cl0939+4713  & 0.41&$0.86^{+0.14}_{-0.10}$&$2.5^{+0.9}_{-0.5}$&0.44&$0.18^{+0.05}_{-0.05}$&$1.2^{+0.4}_{-0.3}$&0.11&$0.09^{+0.02}_{-0.02}$\cr
RXJ1347-1145 & 0.45&$1.14^{+0.07}_{-0.04}$&$6.6^{+0.8}_{-0.7}$&2.20&$0.33^{+0.04}_{-0.04}$&$3.3^{+0.4}_{-0.4}$&0.85&$0.26^{+0.03}_{-0.03}$\cr
3C295        & 0.46&$0.94^{+0.13}_{-0.09}$&$3.8^{+1.1}_{-0.7}$&0.59&$0.16^{+0.03}_{-0.04}$&$1.9^{+0.6}_{-0.3}$&0.21&$0.11^{+0.02}_{-0.03}$\cr
Cl0016+16    & 0.55&$0.98^{+0.05}_{-0.04}$&$5.1^{+0.5}_{-0.4}$&1.31&$0.25^{+0.02}_{-0.02}$&$2.1^{+0.2}_{-0.2}$&0.42&$0.20^{+0.02}_{-0.02}$\cr
MS0451-0305  & 0.55&$1.17^{+0.06}_{-0.08}$&$8.6^{+1.0}_{-1.0}$&1.68&$0.20^{+0.03}_{-0.02}$&$3.8^{+0.4}_{-0.4}$&0.60&$0.16^{+0.02}_{-0.02}$\cr
Cl2236-04    & 0.55&$0.81^{+0.16}_{-0.12}$&$2.9^{+1.2}_{-0.8}$&0.54&$0.19^{+0.07}_{-0.06}$&$1.4^{+0.6}_{-0.4}$&0.19&$0.14^{+0.05}_{-0.04}$\cr
RXJ1716+6708 & 0.81&$0.55^{+0.06}_{-0.03}$&$1.4^{+0.3}_{-0.1}$&0.23&$0.16^{+0.02}_{-0.03}$&$0.7^{+0.2}_{-0.1}$&0.06&$0.09^{+0.01}_{-0.02}$\cr
MS1054-0321  & 0.83&$1.02^{+0.14}_{-0.12}$&$9.4^{+2.4}_{-1.7}$&1.61&$0.17^{+0.04}_{-0.03}$&$3.0^{+0.8}_{-0.5}$&0.46&$0.15^{+0.03}_{-0.03}$\cr
AXJ2019+112  & 1.01&$0.84^{+0.18}_{-0.17}$&$6.9^{+3.4}_{-2.4}$&0.18&$0.03^{+0.01}_{-0.01}$&$3.2^{+1.6}_{-1.1}$&0.93&$0.03^{+0.02}_{-0.01}$\cr
\hline
\end{tabular}
\end{center}
\caption{Total mass, gas mass and gas mass fraction. The first
and the second column give the cluster name and redshift,
respectively. Column (3)
denotes the radius $r_{500}$ which comprises an overdensity of 500 over the
critical density. Columns (4), (5) and (6) list the total mass, the
gas mass and the gas mass fraction within $r_{500}$, respectively. In
columns (7), (8) and (9) the same quantities are listed for a 
radius $0.5\times r_{500}$. The errors listed are only the errors
coming from the uncertainty in the temperature measurement.
}
\label{tab:mass}
\end{table*}

With the assumption of hydrostatic
equilibrium, the integrated total mass can be calculated from the equation
$$
M(r) = {-kr\over \mu m_p G} T \left({ d \ln \rho \over d \ln r }+
                                    { d \ln T    \over d \ln r }\right),
   \eqno(2)
$$
where $\rho$ and $T$ are the density and the temperature of the
intra-cluster gas, and $r$, $k$, $\mu$, $m_p$, and $G$ are the 
radius, the Boltzmann constant, the molecular weight, the proton mass, and
the gravitational constant, respectively. For distant clusters it
is very difficult to measure a temperature gradient. But in any case 
the density gradient is dominating the brackets 
in Eq. 2. Therefore we assume that the
clusters are isothermal with the temperatures listed in
Table~\ref{tab:input}. The resulting masses and gas masses within $r_{500}$
are listed in Table~\ref{tab:mass}. The listed errors on the mass are only
the errors coming from the uncertainty in the temperature. The true
mass uncertainties are larger because of the additional uncertainties
coming from deviations from spherical
symmetry, deviations from hydrostatic equilibrium and projection
effects (about 15\%; see Evrard et al. 1996; Schindler 1996), but these 
are hard to quantify for each cluster individually. Therefore we only
show the errors from the temperature but keep in mind that the true
errors are larger. 

For the luminosity an error of 10\% and for the gas mass an
error of 15\% is used to derive the
error on the power law index. Only for AXJ2019+112 much larger errors,
33\% and 50\%, respectively, are used.
These errors are estimated to include
the uncertainties in the conversion of luminosities in the ROSAT band
to bolometric luminosities, uncertainties in the background
determination and possible contributions of AGNs or background quasars.
The AGN contribution is assumed to be small. For the cluster 3C295, 
which hosts
a luminous radio source and an AGN contribution to the X-ray emission
could be expected, a thorough spatial analysis of the ROSAT/HRI data
was done, and it was concluded that there is no X-ray point source
present in the cluster (Neumann 1999). For the other clusters only
minor contributions are expected as the mean AGN luminosity in the 
ROSAT band is only 
about $5\times 10^{43}$ erg/s (Grupe et al. 1998). Furthermore, 
in none of the spectra  a strong non-thermal component was required to 
fit the data. In Cl0939+472 the X-ray contribution of a  background quasar 
could clearly be identified and was subtracted 
from the luminosity of the cluster (Schindler et al. 1998). 
The contribution to the count rate was less than three 
percent. Therefore, if there are background quasars in other clusters, 
which were not identified, we expect that their contribution is even less 
than three percent. Another possible contamination of the luminosity
can come from cooling flows (Fabian et al. 1994). As will be seen
later this is very obvious for the cluster RXJ1347-1145 with its very
strong cooling flow of $\grsim 3000 \msol$/year (Schindler et
al. 1997). Unfortunately, it is very difficult in distant clusters to
separate cluster emission from cooling flow emission as many
assumptions have to be made and only few observational data are 
available. Therefore we do not make an attempt to correct for the 
cooling flow emission, but exclude the luminosity of RXJ1347-1145 
for the correlation analysis as will be shown later.
For the other clusters in the sample (including 3C295 which has 
a cooling flow 4-9 times weaker than the one in RXJ1347-1145 (see
Neumann 1999))
the cooling flow contribution to the luminosity is probably small 
because for none of them
a deviation from the expected luminosity is visible.

We fit power law functions to all relations for which the linear
correlation coefficient predicts a probability of more than 95\% that
the data are correlated. The fits take into account the errors in both
axes. Note that for the total masses we use only the errors induced by
the uncertainty of the temperature measurement as listed in Table 2. 
Increasing the errors on all data points by a fixed factor on both axes
yields the same best fit results only the errors on the fit parameters 
would be larger. Therefore the
errors on the power law indices are strictly speaking only lower limits. 

Finally, we list some details about particular clusters. For AC118 two
different temperature measurement are published: 9.3 keV (Mushotzky \&
Loewenstein 1997) and 12.1 keV (Mushotzky \& Scharf 1997). 
We use the former value, but assume 
a large error which comprises also the error
range of Mushotzky \& Scharf (1997). For Cl0016+16 two slightly
different temperatures and metallicities are published by Furuzawa et
al. (1998) and Hughes \& Birkinshaw (1998). We use the latter ones. 
For RXJ1347-1145 we use in addition to the data set used in Schindler
et al. (1997) two more data sets observed in the meantime 
resulting in slightly different $\beta$-model parameters.
No analysis of the MS0451-0305 ROSAT/HRI data is published, therefore
we determine countrate and $\beta$-model parameters from the data in
the archive. For two clusters with strong subclusters, 
AC118 and Cl0939+4713, the $\beta$-fit
is centred on the main maximum and the region around the second
maximum is neglected for the fit. As shown in Schindler (1996) this is
a good way to obtain a reliable mass estimate. The $\beta$-model
parameters of MS1054-0321 are not well constrained (Donahue at
al. 1998). We used the central value of their error range for $\beta$ = 0.85
and estimate the error on the core radius to $\pm200$kpc.

\section{Results}

\subsection{Relations of total mass, gas mass and iron mass}

To test whether there is any dependence 
on redshift within our sample we plot various quantities versus
redshift (Fig.~\ref{fig:redshift}).
In this figure, as well as in all other following figures, each
cluster is plotted with a different symbol (see
Table~\ref{tab:legend}). The open symbols (open circle, open triangle,
open square and open hexagon) show the most nearby clusters of the
sample, the
starred and skeletal symbols (star, cross and asterisks) show the
clusters at intermediate redshift and the filled symbols (filled
circle, filled triangle and filled square) indicate the most distant clusters.

\begin{table}[htbp]
\begin{center}
\begin{tabular}{|c|}
\hline
\psfig{figure=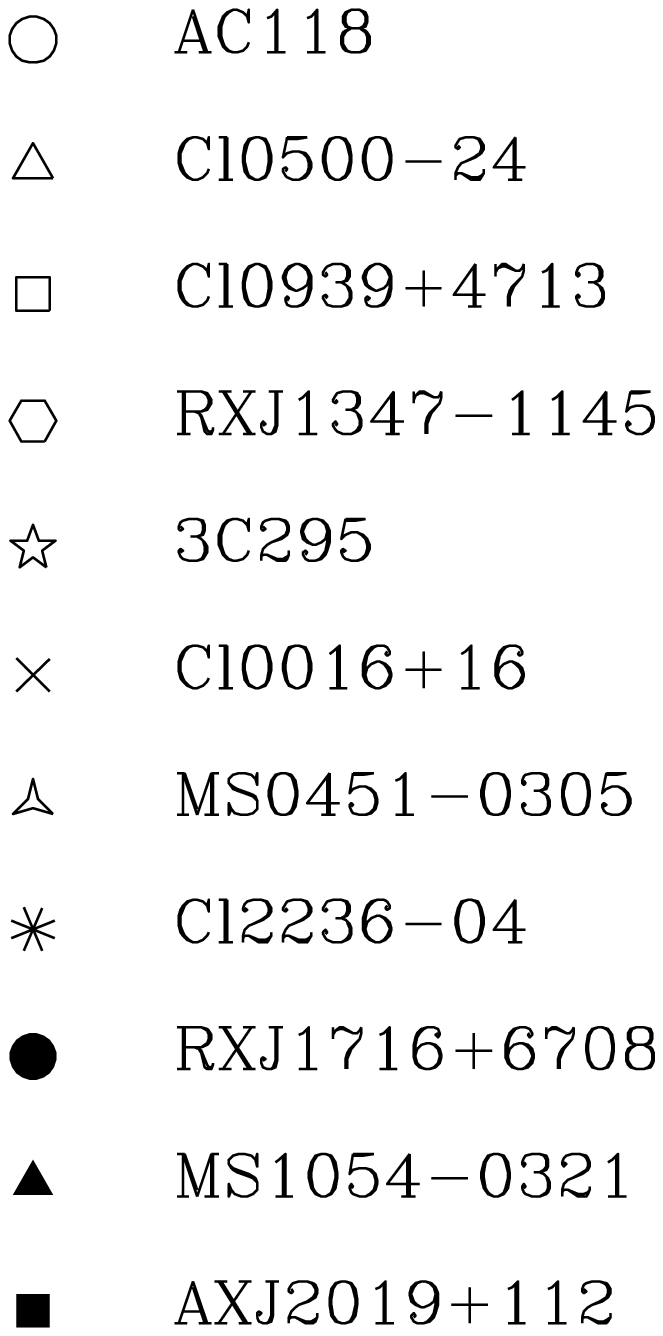,width=4.cm,clip=} \cr
\hline
\end{tabular}
\end{center}
\caption{Symbols used for the various clusters in the figures}
\label{tab:legend}
\end{table}

\begin{figure*} 
\psfig{figure=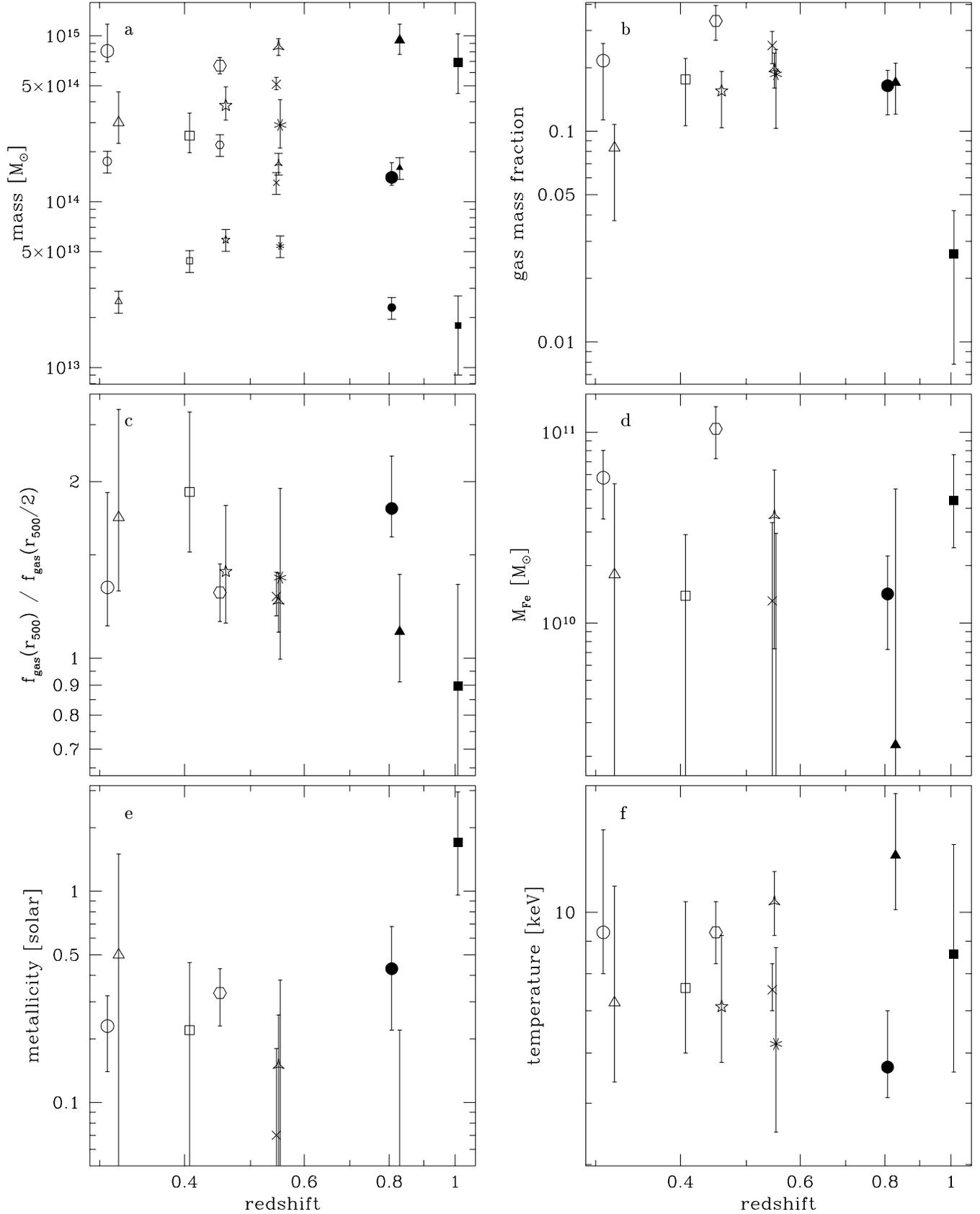,width=17.9cm,clip=} 
%\begin{tabular}{cc}   
%\psfig{figure=z_Mtot.ps,width=8.cm,clip=} \put(-170.,170.){a} &
%\psfig{figure=z_fgas.ps,width=8.cm,clip=} \put(-170.,170.){b} \cr
%\psfig{figure=z_fratio.ps,width=8.cm,clip=}\put(-170.,170.){c} &
%\psfig{figure=z_Fe.ps,width=8.cm,clip=} \put(-170.,170.){d}  \cr
%\psfig{figure=z_m.ps,width=8.cm,clip=}  \put(-170.,190.){e}  &
%\psfig{figure=z_T.ps,width=8.cm,clip=}  \put(-170.,190.){f}  \cr
%%\put(0.,0.){.}
%%\put(80.,800.){b}
%\end{tabular}
\caption[]{Various X-ray properties against redshift.
a)
Gas and total mass. The large symbols
show the total mass, the small symbols show the gas
mass of the corresponding clusters. For explanation of the symbols see
Table 3.
%Table~\ref{tab:leg}.
b) Gas mass fraction,
c) ratio of gas mass fractions at $r_{500}$ and $r_{500}/2$ as a
measure for the relative extent of the gas distribution,
d) iron mass in the intra-cluster gas, e) metallicity, f) temperature.
}
\label{fig:redshift}
\end{figure*}

Fig.~\ref{fig:redshift}a shows the total mass and the gas mass, both
determined within the radius $r_{500}$ (see
Table~\ref{tab:mass}). Both quantities show a considerable scatter, but
no trend with redshift. The total masses range from 
$1.4 \times 10^{14}\msol$ in RXJ1716+6708 to 
$9.4 \times 10^{14}\msol$ in MS1054-0321.  
The gas masses vary between 
$1.8 \times 10^{13}\msol$ (AXJ2019+112) and 
$2.2 \times 10^{14}\msol$ (RXJ1347-1145).

From the total mass and the gas mass we derive the gas
mass fraction $f_{gas}$ at $r_{500}$ (Fig.~\ref{fig:redshift}b). 
Also in the gas mass fraction we see no clear trend with redshift
(only AXJ2019+112 is an obvious outsider).
This result does not confirm the trend of decreasing gas mass fractions by
Tsuru et al. (1997).
The mean value is
$\langle f_{gas} \rangle =0.18$, which is in agreement within the scatter 
with the values for 
nearby samples, e.g. Arnaud \& Evrard (1999): $f_{gas}=0.16$ and 0.20,
respectively, Mohr et al. (1999): $f_{gas}=0.21$ and Ettori \&
Fabian (1999): $f_{gas}=0.17$.   
Therefore we conclude, that we do not see
evolution in the gas mass fraction in these data. 
This result does not confirm the result by  
Ettori \& Fabian (1999) where evolution of the gas mass
fraction is found in a nearby sample.

We see large
variations in the gas mass fraction of more than an order of magnitude 
between individual clusters, ranging from the exceptionally low
fraction of 0.026 in AXJ2019+112 to 0.33
in RXJ1347-1145, very similar to what Ettori \& Fabian (1999) and
Reiprich (1998) found in
nearby samples, but contradicting the result by Wu et al. (1999).

\begin{figure*} 
\psfig{figure=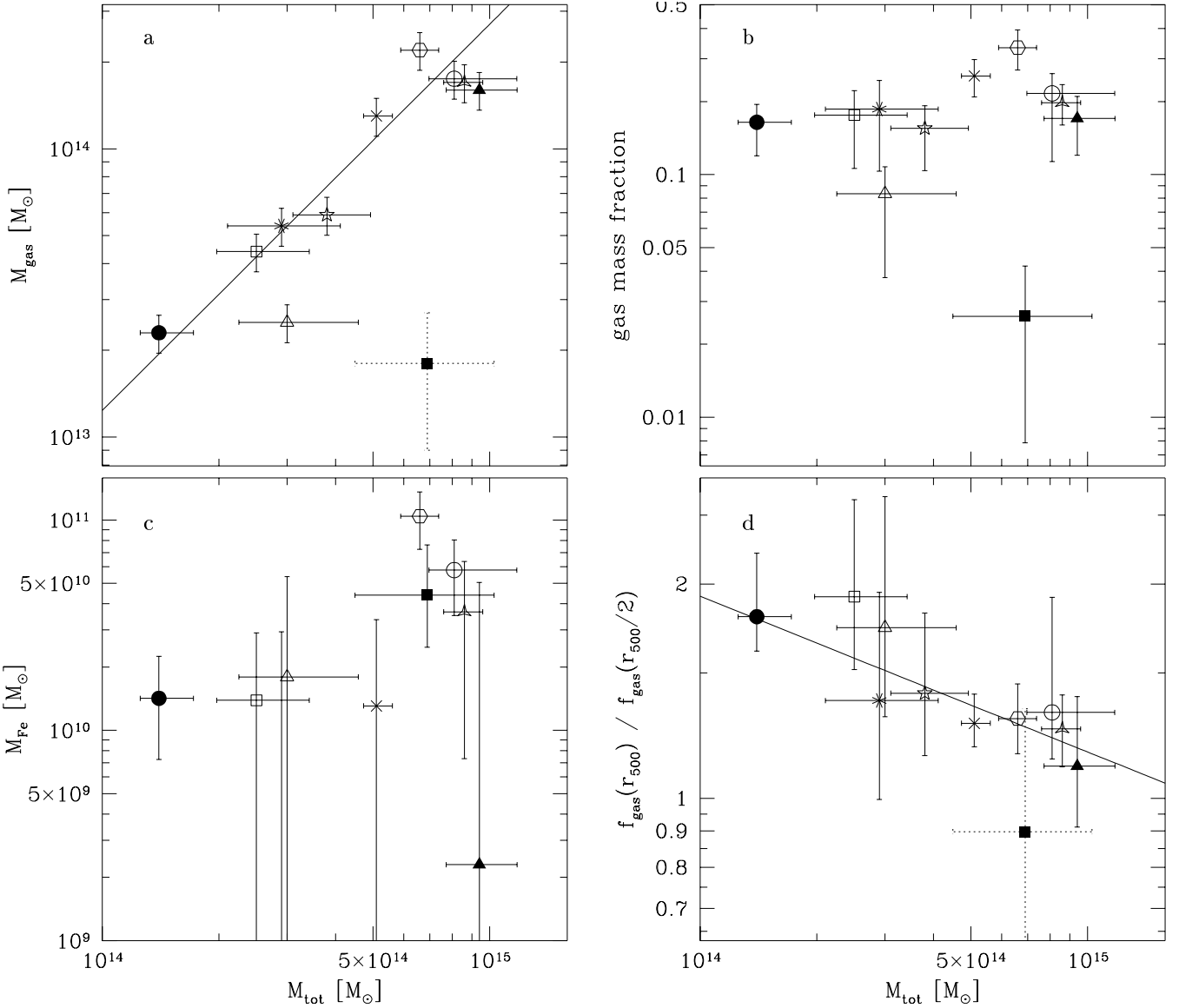,width=17.9cm,clip=}
%\begin{tabular}{cc}   
%\psfig{figure=Mtot_Mgas.ps,width=8.cm,clip=}  \put(-170.,170.){a} &
%\psfig{figure=Mtot_fgas.ps,width=8.cm,clip=} \put(-170.,170.){b} \cr
%\psfig{figure=Mtot_Fe.ps,width=8.cm,clip=}  \put(-170.,190.){c} &
%\psfig{figure=Mtot_fratio.ps,width=8.cm,clip=}  \put(-170.,190.){d} \cr
%\end{tabular}
\caption[]{Various masses versus total mass. a) Gas mass. The solid
line is the best fit excluding AXJ2019+112 (filled square with dashed
error bars). b) The gas mass
fraction is independent of the total mass. c) Iron mass
in the intra-cluster gas. 
d) Ratio of gas mass fractions at $r_{500}$ and
$r_{500}/2$ as measure for the relative extent of the gas
distribution. The solid
line is again the best fit excluding AXJ2019+112.
}
\label{fig:massmass}
\end{figure*}

The gas mass fraction is not constant with radius. We
compare the gas mass fraction at $r_{500}$ with the gas mass fraction
at $0.5\times r_{500}$. The mean gas mass fraction at $0.5\times 
r_{500}$ is 0.13, i.e. smaller than the mean of 0.18 at $r_{500}$. 
In Fig.~\ref{fig:redshift}c we show the ratio of
these fractions $E$ for the individual clusters. We see an
increase of gas mass fraction with radius (i.e. 
$E>1$) in all clusters apart from 
AXJ2019+112. The physical interpretation of $E$ is the extent of the gas
distribution relative to the dark matter extent. 
This means that in general the gas distribution is more
extended than the dark matter, which 
is in agreement with the results for nearby cluster samples by David
et al. (1995), Jones \& Forman (1999), and Ettori \&
Fabian (1999). 
Obviously, cluster evolution is not completely a self-similar process,
but physical processes taking place in the gas must be taken into
account, like e.g. energy input by supernovae, galactic winds or
ram-pressure stripping (see e.g. Metzler \& Evrard 1997; Cavaliere et
al. 1998a). There is no trend of this relative gas extent with  redshift.

Using the gas masses and the metallicities determined from ASCA
observations (see Table~\ref{tab:input}) we can determine
the iron mass in the intra-cluster gas. Unfortunately, the
metallicities have large uncertainties and only for 5 clusters they are
not compatible with zero, which makes it difficult to see any trends
(see Fig.~\ref{fig:redshift}d).
As far as one can see there is no  dependence of the iron mass on redshift.

The lowest two panels of Fig.~\ref{fig:redshift} show the metallicity
and the temperature versus redshift. Also here no
trend is visible within the sample. Within the error bars both
distributions are consistent with a horizontal line. The same result
for the temperature was found for a medium redshift 
sample ($0.14<z<0.54$) by Mushotzky \& Scharf (1997).
Also for the metallicity the  result is in agreement with the result for
medium distant samples (Tsuru et al. 1996; Mushotzky \&
Loewenstein 1997) and with theoretical predictions by Martinelli et
al. (1999). 
The mean value of the metallicity in our sample is
$\langle m \rangle =0.36$. Considering the large error this is in
good agreement with iron abundances of nearby clusters $m=0.2-0.3$
(Fukazawa et al. 1998).

In Fig.~\ref{fig:massmass} we compare different masses with each other.
A trend of an increasing gas mass with total mass which
was found for nearby clusters by Arnaud \& Evrard (1999) is also visible
in our sample when omitting AXJ2019+112 (Fig.~\ref{fig:massmass}a). 
A fit without taking into account AXJ2019+112 yields
$$
M_{gas,500} = 0.12 \, M_{tot,500}^{(1.3\pm0.2)}.
   \eqno(3)
$$
$M_{gas,500}$ and $M_{tot,500}$ are in units of $10^{14} \msol$.
Note that the error on the power law index in Eq. 3 and all following
relations is probably even larger
than indicated here, because we could not take into account all
possible error sources.

Because the exponent in Eq. 3 is close to unity, 
we find basically no  dependence  of the gas mass fraction on the total
mass (see Fig.~\ref{fig:massmass}b). 
Again AXJ2019+112 is lying clearly far away with a gas mass fraction an
order of magnitude too low for its total mass.

In Fig.~\ref{fig:massmass}c the total mass is plotted versus the iron
mass. Again the iron mass is very uncertain because of the large
uncertainties in the metallicity measurements. One can see 
here a marginal trend of
more massive clusters having more iron mass. This can be 
explained easily. The gas mass is proportional to the
total mass ($\approx$ constant gas mass fraction) while the metallicity is
independent of the total mass. As the iron mass is a
product of gas mass and metallicity, it must increase with
increasing total mass. But because of the large errors on the data we
do not attempt a fit through these data.

The relative extent $E$ of the gas distribution with respect to the dark
matter (expressed as the ratio of gas mass fractions at $r_{500}$ 
and $0.5\times r_{500}$) shows an interesting dependence on the total mass
(Fig.~\ref{fig:massmass}d). Clusters with larger masses have smaller
relative gas extents. A fit without taking into
account AXJ2019+112 yields
$$
E = f_{gas}(r_{500}) / f_{gas}(r_{500}/2) = 1.9 \, M_{tot,500}^{(-0.22\pm0.11)},
   \eqno(4)
$$
with $M_{tot,500}$ in units of $10^{14} \msol$. A similar dependence
of the relative gas extent on the total mass
was confirmed by Reiprich (priv. comm.) for
a nearby sample.

\subsection{Mass - temperature relation}

\begin{figure}
\psfig{figure=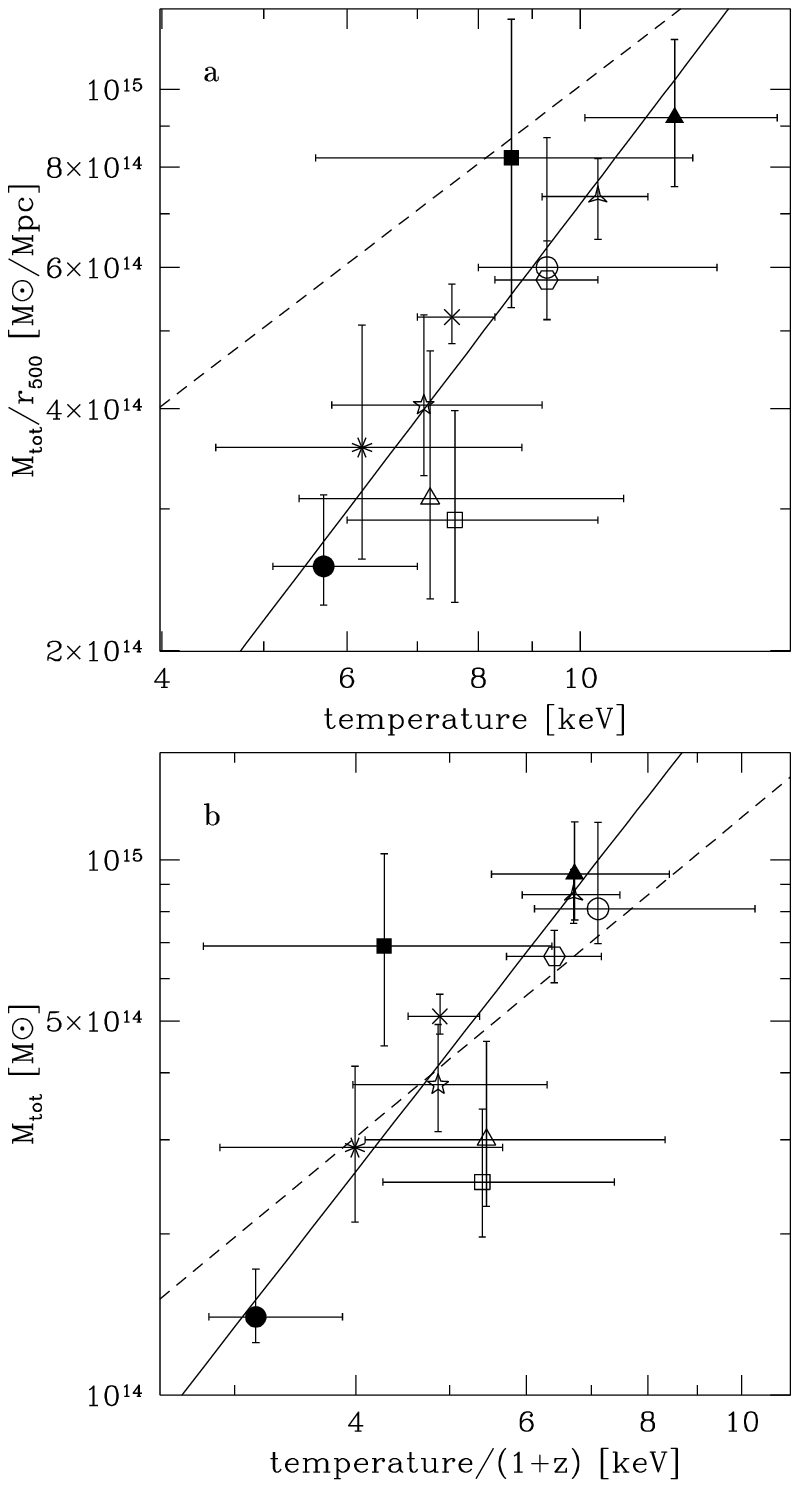,width=8.8cm,clip=}
%\begin{tabular}{c}   
%\psfig{figure=Mr_T.ps,width=8.cm,clip=} \put(-170.,190.){a}  \cr
%\psfig{figure=Mtot_T.ps,width=8.cm,clip=}  \put(-170.,190.){b} \cr
%\end{tabular}
\caption[]{Mass - temperature relations. a) The best fit curve is
shown by the solid line. 
The dashed line is not a fit but the curve expected
from virial considerations with the normalisation of Evrard et
al. (1996). 
b) Equivalent diagram with r being expressed by the definition
of the overdensity contrast. For comparison, the dashed line shows
again the slope expected from virial assumptions, here with an
arbitrary normalisation. 
}
\label{fig:masst}
\end{figure}

Assuming self-similarity and a velocity dispersion proportional to the
X-ray temperature, the viral theorem provides a relation
between total mass, radius and X-ray temperature: $M_{tot,500}/r_{500}
\propto T$. We see this correlation in our data (see
Fig.~\ref{fig:masst}a). But the slope is different from unity which is
the expected 
slope from virial considerations, although the expected slope is 
almost within the error. For comparison, in nearby lensing
observations ($z=0.17-0.54$)
Hjorth et al. (1998) found good agreement with a slope of 1. Our best fit is
$$
M_{tot,500}/r_{500} = 0.14 \, T^{(1.7\pm0.6)},
   \eqno(5)
$$
(compare solid and dashed line in Fig.~\ref{fig:masst}a). $M_{tot,500}$ is
in units of $10^{14} \msol$, $r_{500}$ is in units of Mpc and $T$ is
in units of keV.

Equivalently, $r_{500}$ can be expressed by the definition of the
overdensity $r_{500} \propto M_{tot,500}^{1/3} / (1+z)$ yielding the
relation $M_{tot,500} \propto (T/(1+z))^{3/2}$. 
 For comparison with
nearby clusters we plot also this relation (Fig.~\ref{fig:masst}b). 
We find  a trend of increasing mass with increasing $T/(1+z)$.
A fit taking into account the errors in
the temperature and the error in the mass yields the following relation
$$
M_{tot,500} =  0.10 \, \left( {T \over 1+z} \right) ^{(2.3\pm0.8)},
   \eqno(6)
$$
shown in Fig.~\ref{fig:masst}b as solid line. $M_{tot,500}$ is again
in units of $10^{14} \msol$ and $T$ in units of keV.
The slope is again steeper than the 1.5 expected from the virial
theorem. For comparison we plot also a line with this slope in
Fig.~\ref{fig:masst}b. Obviously, the errors and the scatter are so
large that no final conclusions can be drawn as the virial value
is included in the errors.  
Also in nearby clusters slopes larger than 1.5 were found. Ettori \&
Fabian (1999) find in a sample of cluster with redshifts between 0.05
and 0.44 an exponent of $1.93\pm0.09$. Horner et al. (1999) find
an exponent of 1.8 -- 2.0 and Reiprich (1998) finds an exponent of 2.0. 

\begin{figure}
\psfig{figure=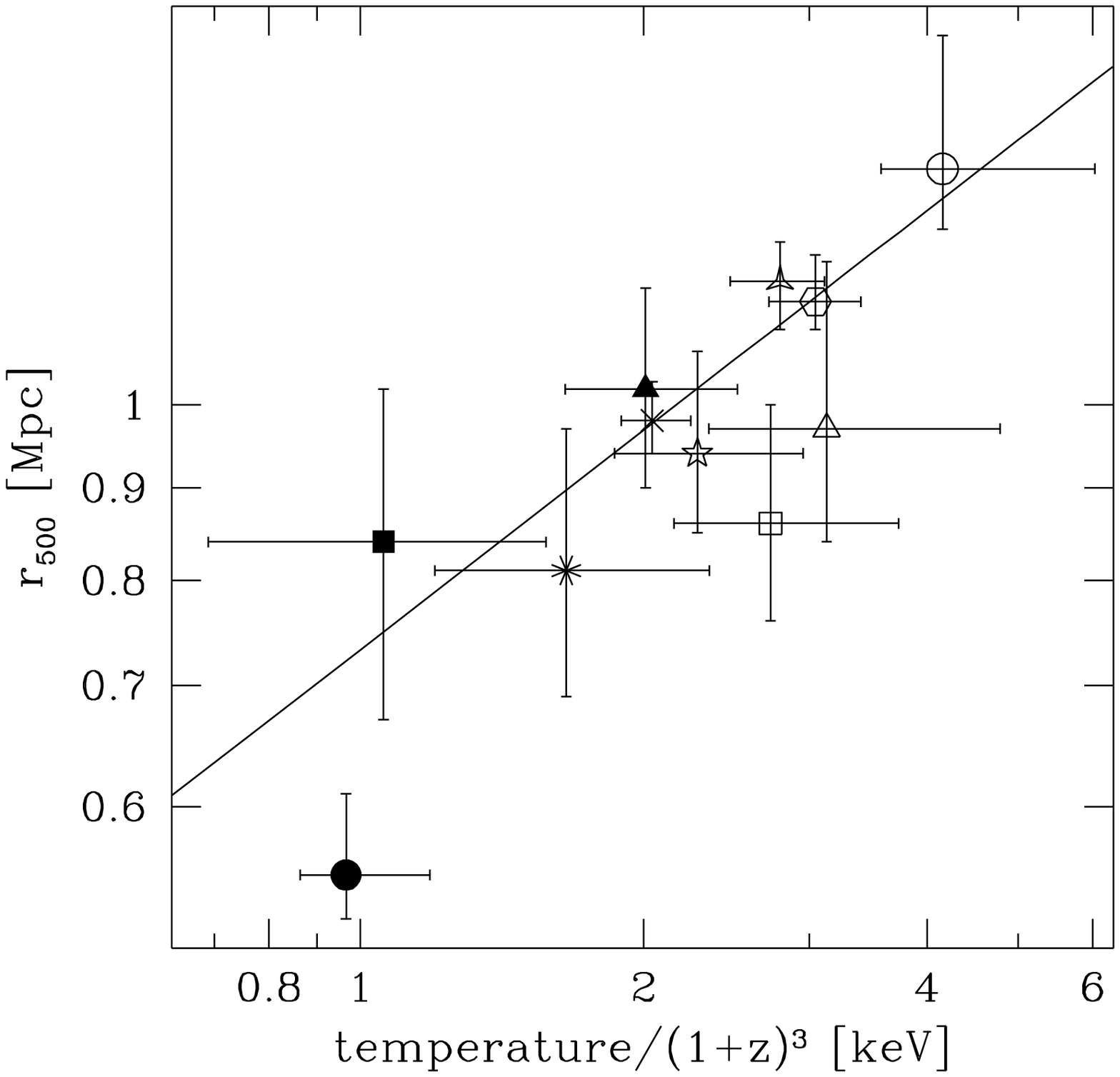,width=8.cm,clip=} 
\caption[]{Temperature -- $r_{500}$ relation.
}
\label{fig:rt}
\end{figure}

Another way of putting the same relation is relating the radius to
the temperature: $r_{500} \propto T^{1/2} / (1+z)^{3/2}$. Of course, the
relation is also visible in this way (see Fig.~\ref{fig:rt}). The fit
gives a slightly smaller slope than the expected 0.5, but 0.5 is well
included in the error:
$$
r_{500} = 0.73 \, \left( {T \over (1+z)^3} \right) ^{(0.40\pm0.24)},
   \eqno(7)
$$
with $r_{500}$ in Mpc and T in keV. Mohr \& Evrard (1997) found also
a radius -- temperature relation for nearby clusters. But different
from $r_{500}$ they used the isophotal radius of the X-ray emission and
found a larger slope of 0.93.

\begin{figure*} 
\psfig{figure=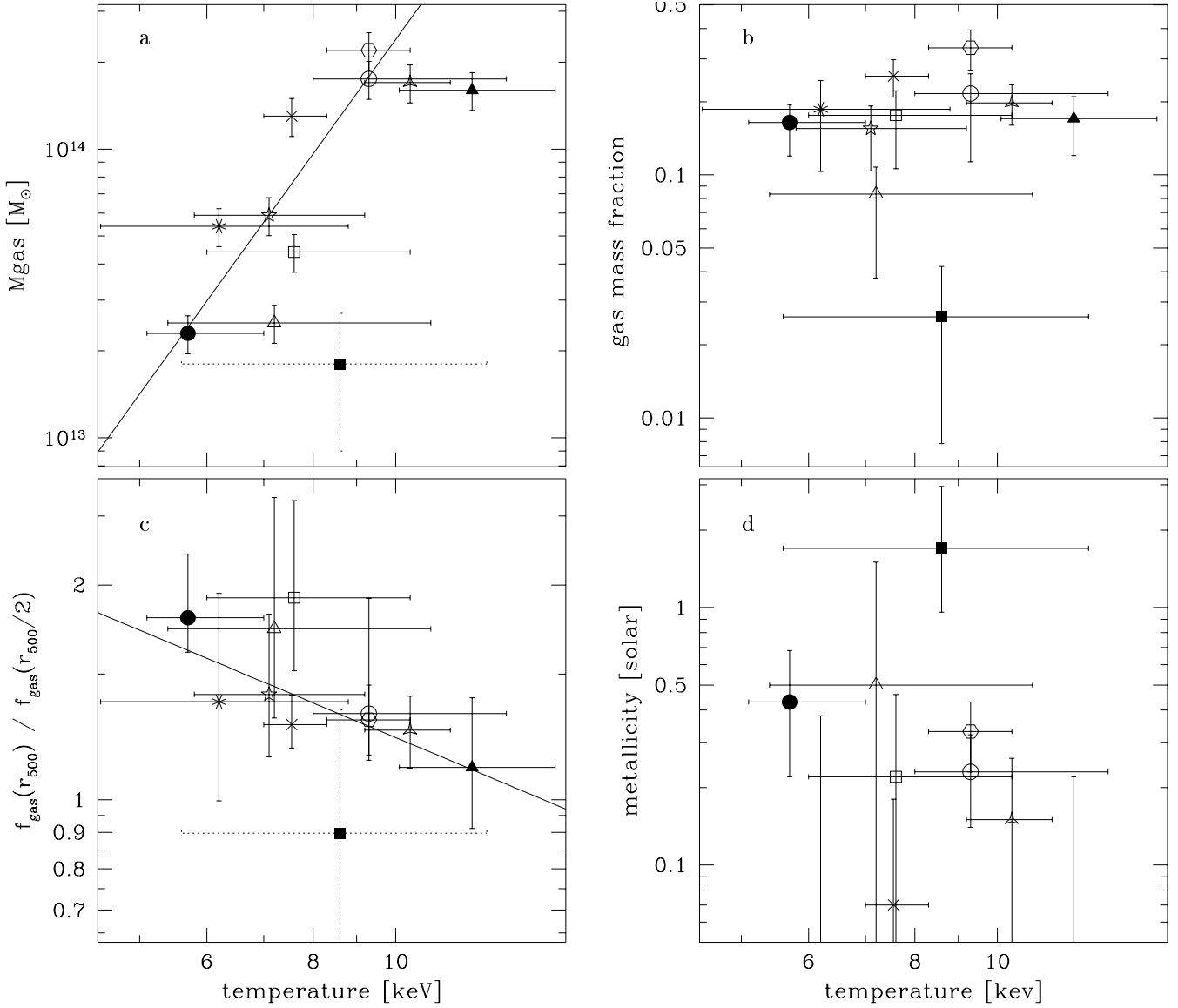,width=17.9cm,clip=} 
%\begin{tabular}{cc}   
%\psfig{figure=T_Mgas.ps,width=8.cm,clip=}  \put(-170.,170.){a} &
%\psfig{figure=T_fgas.ps,width=8.cm,clip=} \put(-170.,170.){b}  \cr
%\psfig{figure=T_fratio.ps,width=8.cm,clip=} \put(-170.,190.){c}  &
%\psfig{figure=T_m.ps,width=8.cm,clip=}  \put(-170.,190.){d} \cr
%\end{tabular}
\caption[]{a) 
b) Various quantities versus the temperature. a) Gas mass, b) gas mass
fraction, c) ratio of gas mass fractions at $r_{500}$ and
$r_{500}/2$), d) metallicity. AXJ2019+112 (solid square) is excluded
in the fits.
}
\label{fig:gast}
\end{figure*}

Furthermore, we test the relation of the gas mass and the gas mass
fraction on the temperature. As expected from the gas mass --
total mass relation, there is also a relation between
gas mass and temperature (see Fig.~\ref{fig:gast}a), in particular  
when the gas mass outsider AXJ2019+112 is removed. A fit without
AXJ2019+112 yields
$$
M_{gas,500} = 2.0\times10^{-4} \, T^{(4.1\pm1.5)},
   \eqno(8)
$$
with $M_{gas,500}$ in units of $10^{14}\msol$ and T in keV. This
correlation was also found in nearby clusters (Reiprich 1998; Jones \&
Forman 1999). For comparison, Reiprich (1998) finds an exponent of 2.9.

As expected from the non-correlation of
gas mass fraction with the total mass, we find also no correlation
between the gas mass fraction and the temperature (see
Fig.~\ref{fig:gast}b). This result is in good agreement
with Mohr et al. (1999). They find a mild dependence comparing low temperature 
clusters ($T<5$keV) with high temperature clusters ($T>5$keV). For the
high temperature clusters alone, in which category all our clusters
fall, they find no dependence.

We find an interesting correlation between the relative gas extent and
the temperature (Fig.~\ref{fig:gast}c), which is of course related to
the dependence of the relative extent on the total mass, shown above.
The extent of the gas relative to the extent of the dark matter tends
to be larger in lower temperature clusters. Excluding again
AXJ2019+112 one finds
$$
E = f_{gas}(r_{500})/ f_{gas}(r_{500}/2) = 3.9 \, T^{(-0.50\pm0.34)},
   \eqno(9)
$$ 
with the temperature in keV.

We see no correlation between temperature and 
metallicity in our sample (Fig.~\ref{fig:gast}d) confirming the result
by Tsuru et al. (1997).

\begin{figure}
\psfig{figure=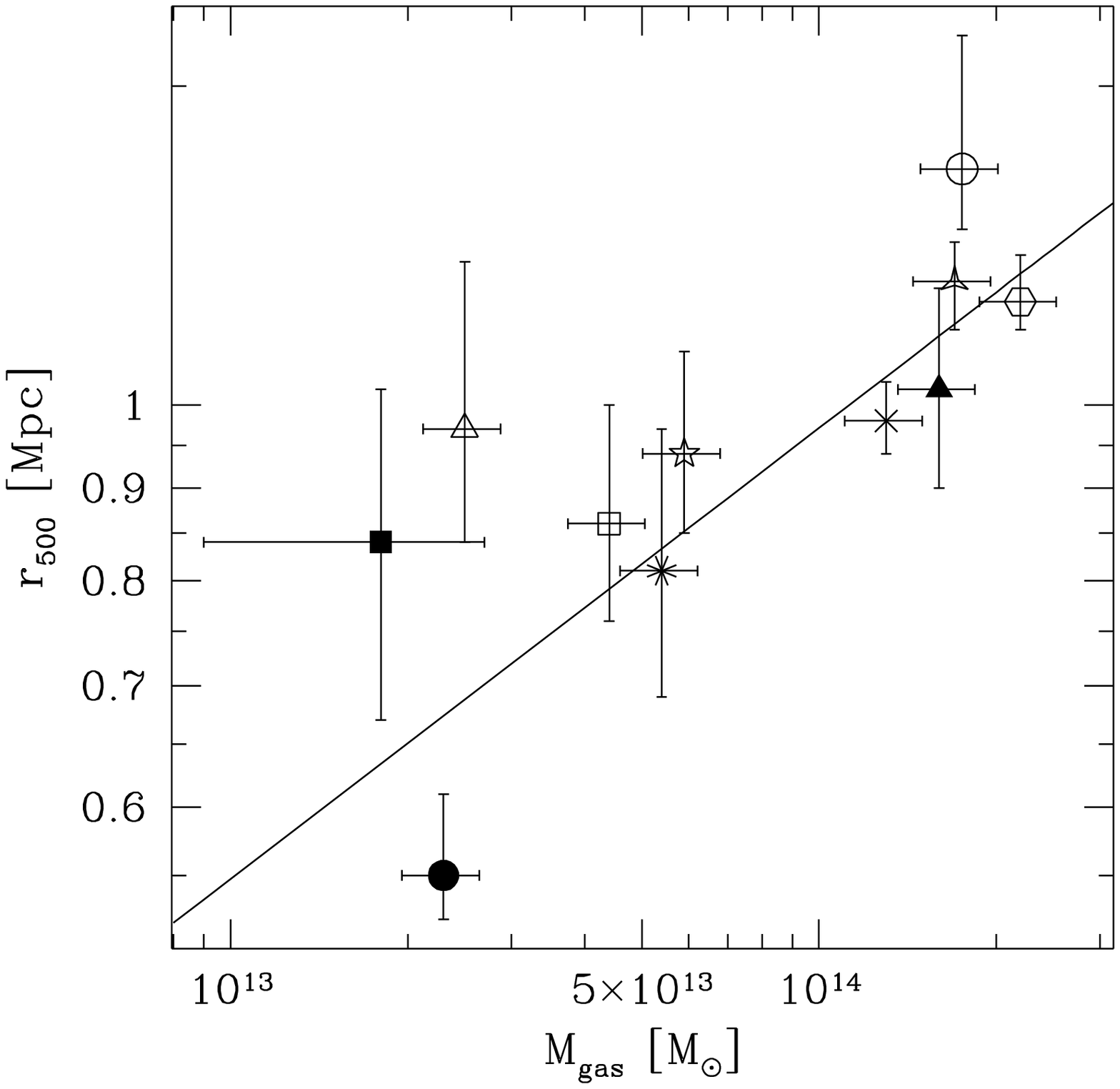,width=8.cm,clip=} 
\caption[]{Gas mass -- $r_{500}$ relation.
}
\label{fig:gasr}
\end{figure}

As the definition of the overdensity contains a relation between total
mass and radius, and there is also a relation between total mass and
gas mass, we expect a relation between the gas mass  and  $r_{500}$,
which is indeed visible in Fig.~\ref{fig:gasr}. A fit yields
$$
r_{500} = 0.97 \, M_{gas,500}^{(0.25\pm0.04)},
   \eqno(10)
$$
with $r_{500}$ in units of Mpc and $M_{gas,500}$ in $10^{14}\msol$.

\subsection{Relations with the X-ray luminosity }

\begin{figure}
\psfig{figure=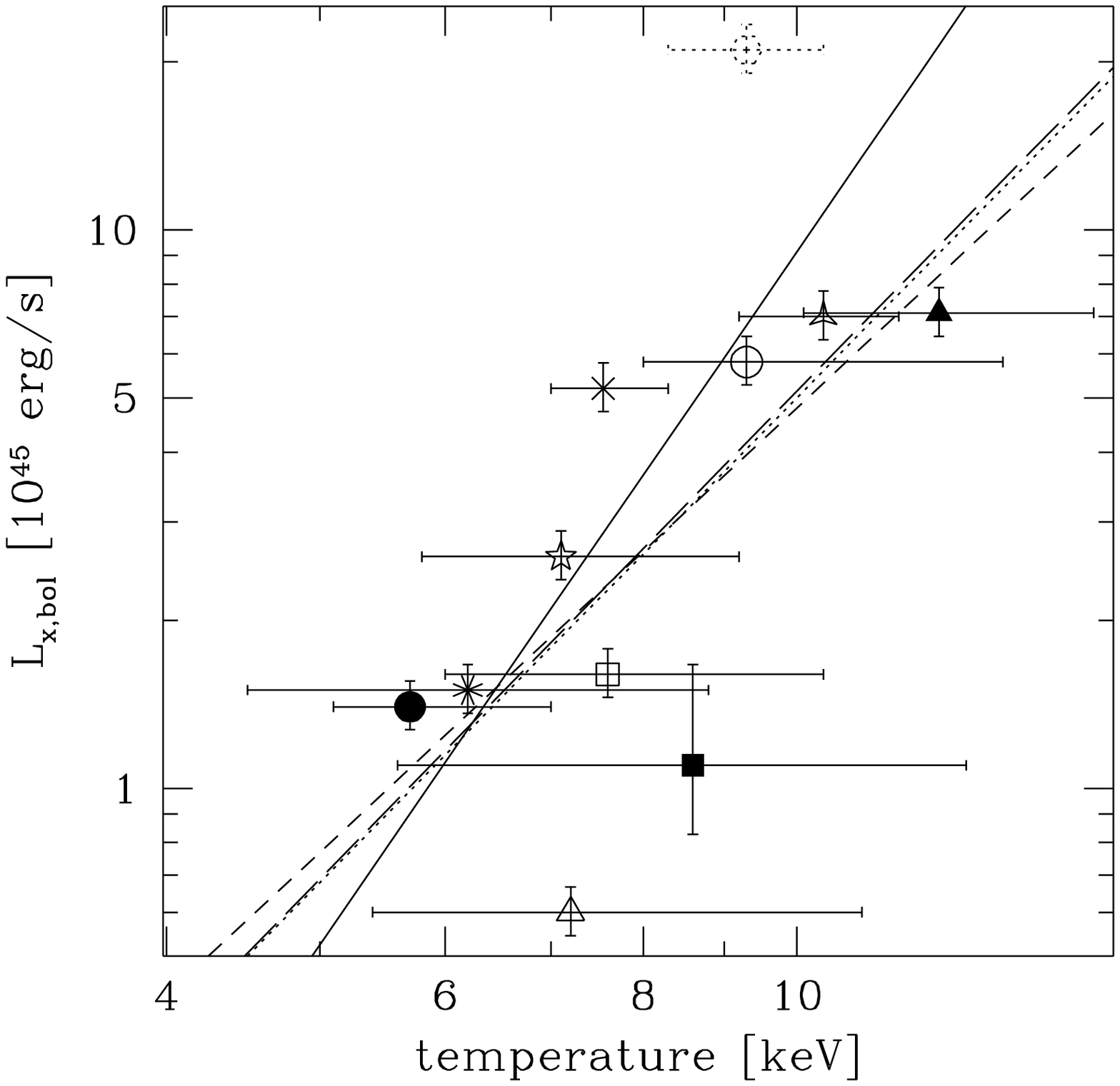,width=8.cm,clip=} 
\caption[]{Luminosity -- temperature relation. The best fit which
excludes  RXJ1347-1145 (dotted hexagon) is indicated by a solid line.
Some $L_X - T$ relations for nearby
clusters are also shown: 
Arnaud \& Evrard 1999 (dotted line), Allen \& Fabian 1998
(long-dashed line) and Markevitch
1998 (short-dashed line).

Apart from RXJ1347-1145
(dotted hexagon), which lies well above the other clusters because of the
strong cooling flow, all other clusters are consistent with $L_X - T$ 
relations for nearby
clusters: Arnaud \& Evrard 1999 (dotted line), Allen \& Fabian 1998
(long-dashed line) and Markevitch
1998 (short-dashed line). Our fit, excluding RXJ1347-1145 (dotted hexagon),
is indicated by a solid line. 
}
\label{fig:lxt}
\end{figure}

The relation between the X-ray luminosity and the temperature, two
directly observable quantities, is
often used to compare gas and dark matter distribution in clusters, 
as the X-ray luminosity is connected with the gas while the
temperature is related to the total mass of the cluster.
Also we find in our data  
the well-known luminosity -- temperature relation (see
Fig.~\ref{fig:lxt}). RXJ1347-1145 is lying above the other clusters,
which is expected because of its very strong cooling flow (Fabian et
al. 1994). Therefore it is excluded for the fit
$$
L_{X,bol} = 6.9\times 10^{-4} \, T ^{(4.1\pm1.7)},
   \eqno(11)
$$
with $L_X$ in units of $10^{45}$ erg/s and T in keV. We find a larger
slope than recently derived $L_X - T$ relations for nearby
clusters (Arnaud \& Evrard 1999: 2.9; Allen \& Fabian 1998: 2.9;
Markevitch 1998: 2.6; Jones \& Forman 1999: 2.8) 
and for a sample including distant cluster but being
heavily dominated by nearby clusters (Wu et al. 1999: 2.7), 
but the slopes for nearby clusters 
are still within the error. 
Some  nearby $L_X - T$ relations are also shown in Fig.~\ref{fig:lxt}
for comparison.
Introducing a term $(1+z)^A$ (see Oukbir \& Blanchard 1997) 
does not reduce the scatter of the data and does not improve the
agreement with the nearby relations. This can be seen directly in
Fig.~\ref{fig:lxt} where the term $(1+z)^A$ would shift the starred
symbols slightly and the filled symbols considerably to the right. As
these symbols lie on both sides of the fit curve, it would not help to
reduce the scatter.
Our conclusion from this is that we cannot see significant signs of
evolution in the $L_X - T$ relation. The same conclusion was also
drawn from medium distant samples by Tsuru et al. (1996) and
Mushotzky \& Scharf (1997) %($0.14 < z < 0.54$)
and from 
samples covering all redshifts up to $z=1$  by Tsuru et al. (1997) and
Sadat et al. (1998).

\begin{figure*}
\psfig{figure=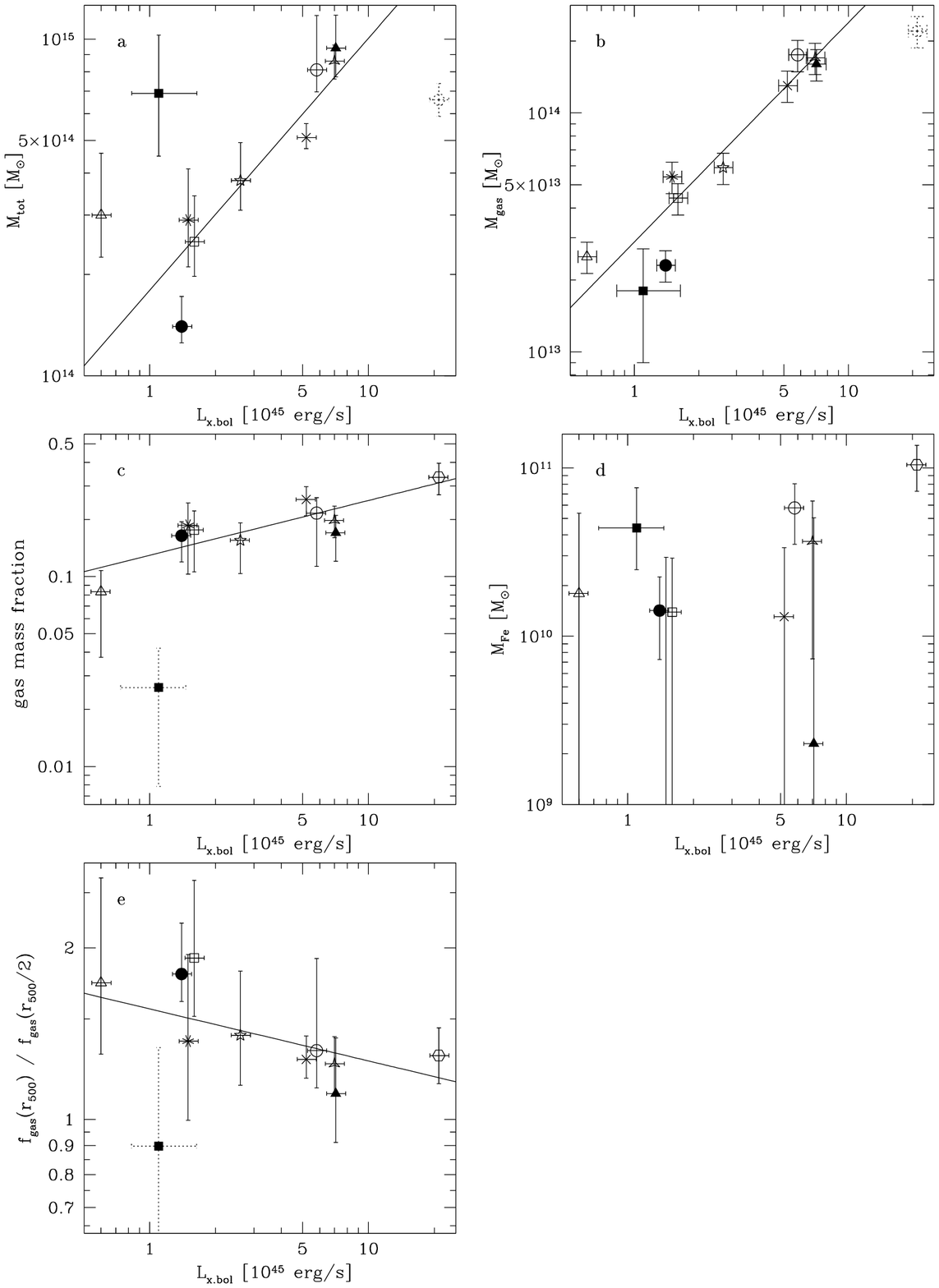,width=17.9cm,clip=} 
%\begin{tabular}{cc} 
%\psfig{figure=lx_Mtot.ps,width=8.cm,clip=} \put(-170.,190.){a}  &
%\psfig{figure=lx_Mgas.ps,width=8.cm,clip=} \put(-170.,190.){b}  \cr
%\psfig{figure=lx_fgas.ps,width=8.cm,clip=}  \put(-170.,190.){c} &
%\psfig{figure=lx_Fe.ps,width=8.cm,clip=}  \put(-170.,190.){d} \cr
%\psfig{figure=lx_fratio.ps,width=8.cm,clip=}  \put(-170.,190.){e} &
%\end{tabular}
\caption[]{Luminosity versus total mass (a), gas mass (b), gas mass
fraction (c), iron mass (d) and the relative gas extent (e) expressed
as the ratio gas mass fraction at $r_{500}$ and $0.5\times r_{500}$. The
fits in (a) and (b) exclude RXJ1347-1145 (dotted hexagon) because of
its strong cooling flow. In the fits of (c) and (e) the gas  mass
outlier AXJ2019+112 is excluded. 
}
\label{fig:lxm}
\end{figure*}

As expected from the correlations between mass--temperature and
luminosity--temperature, 
the luminosity and the total mass show a correlation as well (see
Fig.~\ref{fig:lxm}a).
A fit with a power law (again excluding RXJ1347-1145) yields
$$
M_{tot,500} = 1.8 \, L_{X,bol}^{(0.75\pm0.10)},
   \eqno(12)
$$
($L_{X,bol}$ in units of $10^{45}$ erg/s, $M_{tot,500}$ in units of $10^{14}
\msol$). For comparison, in a sample of 106 nearby clusters an
exponent of 0.8 was found (Reiprich \& B\"ohringer 1999).

The luminosity shows a tight correlation with the gas mass (see
Fig.~\ref{fig:lxm}b) 
$$
M_{gas,500} = 0.29 \, L_{X,bol}^{(0.92\pm0.07)},
   \eqno(13)
$$
($L_{X,bol}$ in units of $10^{45}$ erg/s, $M_{gas,500}$ in units of $10^{14}
\msol$, RXJ1347-1145 excluded). 
As the X-ray emission is proportional to the square of
the gas density this relation gives hints on the gas distribution. 
The small scatter in the
above relation shows that the gas in these clusters has similar
distributions, e.g. not different degrees of clumpiness.

This tight correlation of gas mass and luminosity was also found in a
nearby sample by Jones \& Forman (1999) although they measured the
mass with a fixed radius of 1 Mpc.
Our result is also in relative good 
agreement with the result for a nearby sample by Reiprich (1998) who used the
luminosity in the ROSAT band (0.1-2.4 keV) and found a slope of 1.08.
Cooray (1999) finds a different slope of $0.66\pm0.06$ using the
luminosity in the 2-10 keV band. The reason for the difference is
probably that gas masses within a fixed radius of 0.5 Mpc were used in
the analysis by Cooray.

The gas mass fraction shows a marginal trend to increase with
luminosity (see Fig.~\ref{fig:lxm}c, again the gas mass fraction
outlier AXJ2019+112 is excluded). This trend is mainly caused by the
very low and very high luminosity clusters Cl0500-24 and
RXJ1347-1145. The other clusters would be perfectly consistent with a
horizontal line. As the linear correlation coefficient predicts a high
probability for a correlation between the luminosity and the gas mass
fraction, we try to fit these data, but the large
error on the slope reflects the consistency with a constant gas mass
fraction  
$$
f_{gas,500}=M_{gas,500}/M_{tot,500} = 0.13 \, L_{X,bol}^{(0.29\pm0.86)},
   \eqno(14)
$$
(with $L_{X,bol}$ in units of $10^{45}$ erg/s). For comparison, in a 
nearby sample with a much larger range of luminosities Jones \& Forman
(1999) find an increase of the gas mass with luminosity.

Fig.~\ref{fig:lxm}d shows the iron mass versus luminosity. There might
be a marginal trend of increasing iron mass with luminosity, but again the
error bars on the iron mass are too large to get any significant result.

The relative gas extent $E$ (Fig.~\ref{fig:lxm}e) -- expressed as the
ratio of gas mass 
fraction at $r_{500}$ and $0.5\times r_{500}$ -- shows a less
pronounced correlation with luminosity than with total mass
(Fig.~\ref{fig:massmass}d) or
with temperature (Fig.~\ref{fig:gast}c). A fit excluding AXJ2019+112
yields 
$$
E = f_{gas}(r_{500})/ f_{gas}(r_{500/2}) = 87 \, L_{X,bol}^{(-0.091\pm0.063)},
   \eqno(15)
$$
with $L_{X,bol}$ in units of $10^{45}$ erg/s. 

\subsection{$\beta$-model parameters}

\begin{figure}
\psfig{figure=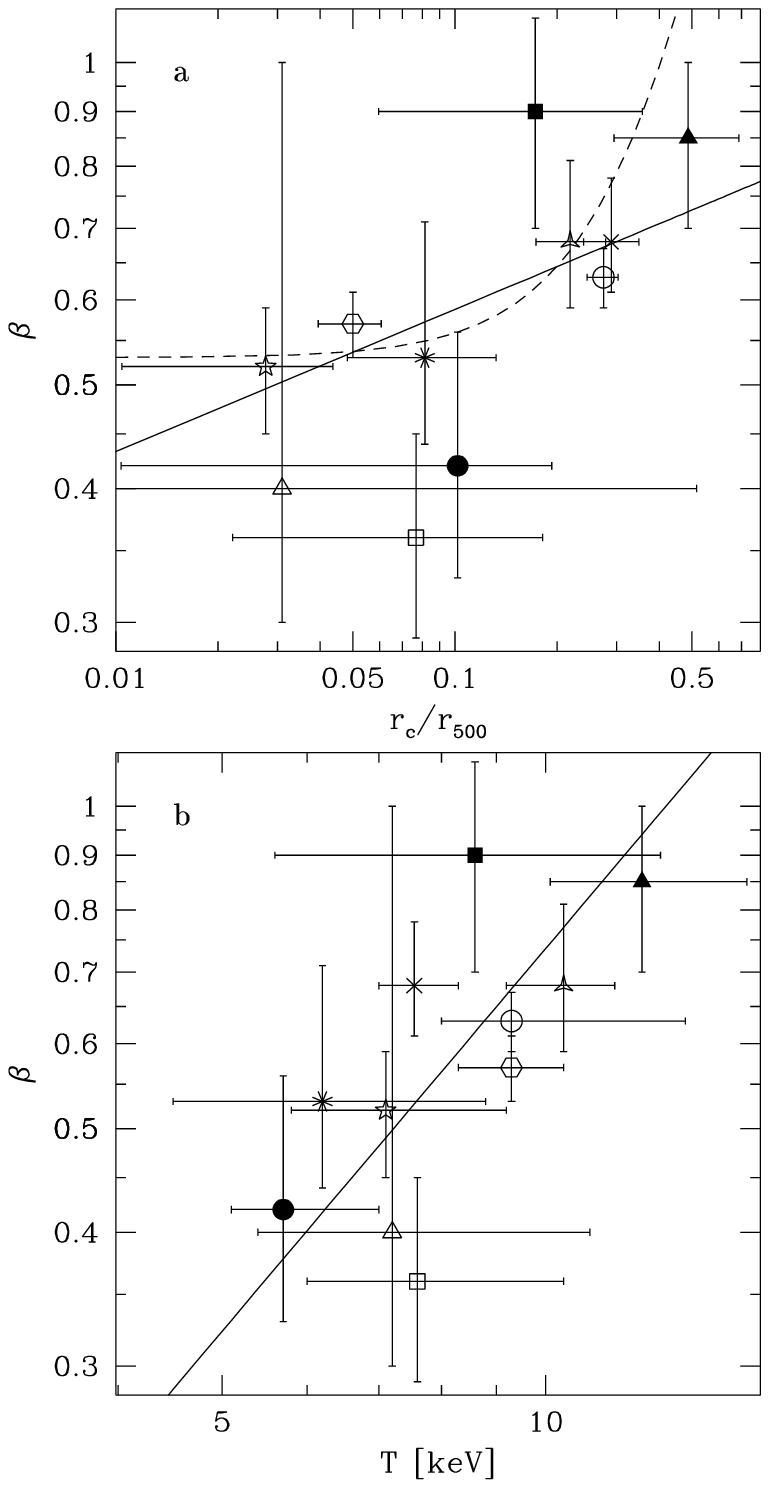,width=8.8cm,clip=}
%\begin{tabular}{c} 
%\psfig{figure=rc_b.ps,width=8.cm,clip=} \put(-170.,190.){a} \cr
%\psfig{figure=T_b.ps,width=8.cm,clip=} \put(-170.,190.){b} \cr
%\end{tabular}
\caption[]{a) Core radius -- $\beta$ relation. The dashed line is the best
fit core radius -- $\beta$ relation by Neumann \& Arnaud (1999). The
full line is a fit with a power law taking into account the errors
in both parameters. b) Temperature -- $\beta$ relation.
}
\label{fig:rcb}
\end{figure}

Neumann \& Arnaud (1999) found a correlation between the fit
parameters of the $\beta$ model
-- the core radius $r_c$ and the slope $\beta$ -- when normalising the
core radius. They normalised their core radii to $r_{200}$. As
the X-ray emission of distant clusters cannot be traced out to such
large radii we normalised the core
radii to $r_{500}$ (see Fig.~\ref{fig:rcb}a). 

Obviously, there is also a correlation between the core radius $r_c$
and the slope $\beta$ in distant clusters. The error
bars in Fig.~\ref{fig:rcb}a  seem large 
and are somewhat misleading. The
error range is not the whole rectangle defined by these error bars,
but more like a very elongated ellipse
from the lower left corner to the upper right corner.

A fit with a power law taking into account the errors in both
parameters yields
$$
\beta = 0.80 \, \left( {r_c \over r_{500}} \right)^{(0.13\pm0.04)}.
   \eqno(16)
$$
For comparison we plot in Fig.~\ref{fig:rcb}a also the fit curve by Neumann \&
Arnaud (1999) for nearby clusters. 
after converting their normalising
radius with a factor 0.63 to ours.
Due to the large error bars we
cannot determine which fit function is the better one.
Both come close to all clusters with small error bars and seem to fit the data
equally well.

In Fig.~\ref{fig:rcb}b we show the dependence of $\beta$ on the
temperature. There is a trend to find larger $\beta$ values for
larger temperatures as was also found for more nearby clusters
(Mohr \& Evrard 1997; Arnaud \& Evrard 1999; Jones \& Forman 1999). We find
$$
\beta = 0.048 \, T ^{(1.2\pm0.5)},
   \eqno(17)
$$
with T in keV.

\section{Summary and Conclusions}

Using ROSAT and ASCA results of distant clusters we derive bolometric
luminosities, total masses, gas masses, gas mass fractions and iron
masses. We compare these with directly measured quantities:
temperature, metallicity, redshift, core
radius and the slope parameter $\beta$.
We find clear positive correlations between the following quantities:
\begin{itemize}
\item{} total mass
\item{} gas mass
\item{} temperature
\item{} luminosity
\item{} $r_{500}$
\item{} $\beta$
\item{} normalised core radius.
\end{itemize}
and a negative correlation of the
\begin{itemize}
\item{} relative gas extent
\end{itemize}
with the other quantities. No correlations are found between the gas
mass fraction and the other quantities.

All relations are in agreement within the errors with relations found
in nearby cluster samples, i.e. consistent with no evolution, although
we find for some relations slightly different slopes, but the
differences are not significant because of the large
uncertainties. Furthermore, we find no trend with redshift in
the quantities themselves. 
 
A low $\Omega$ is required to explain the high gas mass
fraction  of $\langle fgas \rangle = 0.18$, because in an
$\Omega=1$ universe this value is much higher than
the baryon fraction predicted by primordial nucleosynthesis.
A low $\Omega$ is also favoured by the fact that
we see no evolution in temperature or mass with redshift, because in
$\Omega=1$ universe one would expect on average cooler clusters at
high redshift (Oukbir \& Blanchard 1992). Furthermore, as the changes in
the luminosity temperature relation are expected to be larger in an high
$\Omega$ universe (Eke et al. 1998), our results are consistent also
here with low $\Omega$, but $\Omega=1$ can not be excluded because of
the large uncertainties. Summarising, these quantities point towards a
low $\Omega$ universe, but for final conclusions 
they need to be measured with higher accuracy. 

While some relations have been known for a long time from nearby samples,
e.g. the luminosity-temperature relation (Mushotzky 1984) 
or the increase of the gas
mass fraction with radius, others are not well known yet: 
a negative correlation of
the relative gas extent and cluster mass or temperature and a tight
correlation between the bolometric luminosity and gas mass.
Both give interesting insights into cluster structure and formation.

The relative gas extent $E$, defined as the ratio of the gas mass fraction
at $r_{500}$ and $0.5\times r_{500}$, is a measure of how fast the gas
mass fraction is increasing with radius and as such a measure of how 
extended the gas distribution is with respect to the dark matter
distribution. Large values of the ratio $E$ -- corresponding to very
extended gas distribution -- are found in the clusters with low mass
and low temperature, while low ratios $E$ of almost unity -- corresponding
to similar distributions in gas and dark matter -- are found in
massive and hot clusters. This can have different reasons.
One possibility is that it hints to physical processes in the gas,
which are assumed to be responsible for the increase of gas mass
fraction with radius (e.g. Metzler \& Evrard 1997; Cavaliere et
al. 1998a). If gas is placed artificially into a model cluster potential in
hydrostatic equilibrium the distributions of gas and dark matter
have the same slope at radii larger than the core radius (Navarro et
al. 1996), therefore one would expect a priori a ratio $E\approx1$. 
It might be that this additional heat input affects low
mass clusters more that massive clusters, so that a massive cluster can
maintain a ratio $E=1$ while in the smaller clusters the gas is
becoming more and more extended.
Another possibility is that this relation points to some other hidden
correlation. For example, if the clusters have temperature profiles
declining with radius, with our assumption of isothermality we are 
underestimating $E$. If e.g. the temperature profiles in
less massive cluster were steeper than in more massive clusters we
would also expect a gradient in $E$ with mass. This is just an example
to show how different correlations could be connected. In this case it
is probably not an effect of temperature profiles, because Markevitch
et al. (1998) found the same temperature gradient in hot and cool clusters.
This dependence of the relative gas extent $E$ should be tested in
nearby clusters, where the masses can be measured with higher accuracy.

We find large variations in the gas mass fraction, similar to the
results of nearby clusters (Reiprich 1998; 
Ettori \& Fabian 1999). These large
variations, which span an order of magnitude, have some implications on
cluster formation. If all the clusters had originally the same small
gas mass fraction and all the differences came later by different
amounts of gas released by the cluster galaxies, larger metallicities
in clusters with high gas mass fraction would be expected. But this is
not observed. Therefore the difference must be caused at least
partially by the primordial distribution of baryonic and non-baryonic
matter. 

The tight correlation of gas mass and bolometric luminosity gives a
hint that there is not a lot of clumpiness on small scales in the
intra-cluster gas. If there would be
different degrees of clumpiness one would expect a large scatter
around the fit, because with the same amount of gas a clumpier medium
produces more photons as the emission is proportional to
the square of the density. Although a non-clumpy medium 
is implicitly assumed for the
calculation of the gas mass, the fact that everything is consistent is
an interesting result. It confirms results found by the
comparison of X-ray data and Sunyaev-Zel'dovich measurements. As the two
measurements have different dependencies on the gas density a large
degree of clumpiness could be ruled out (e.g. Holzapfel et al. 1997).

Unfortunately, the errors in the metallicity measurement are very
large (only half of clusters in the samples have metallicities
non-consistent with zero). Hence, also the errors in the iron masses
are very large, so that no definite conclusions about the metal
injection into the intra-cluster medium and its time dependence 
can be drawn. 

The comparison with other clusters shows that AXJ2019+112 is really an
exceptional cluster. It is not only the most distant cluster in the
sample, but it is also the only cluster found indirectly by the
gravitational lensing effect. It has by far the highest metallicity, 
the lowest gas mass fraction and the lowest relative gas
extent. Unfortunately, the morphological parameters could not be very
well constrained from the ROSAT/HRI observation because only about 80
source counts were detected (Hattori et al. 1997). Therefore, we test
what effects a small change of the morphological parameters would have
on the masses. Assuming the slope $\beta$ was 0.6
(instead of the best fit value $\beta=0.9$), one finds a total mass of
$M_{tot,\beta=0.6}=3.6\times10^{14}\msol$ and a gas mass of 
$M_{gas,\beta=0.6}=0.32\times10^{14}\msol$.
Comparing these with the masses for  $\beta=0.9$,
$M_{tot,\beta=0.9}=6.9\times10^{14}\msol$ and 
$M_{gas,\beta=0.9}=0.18\times10^{14}\msol$, and with 
the masses expected from the luminosity -- mass relations,
$M_{tot,exp}=2.0\times10^{14}\msol$ and 
$M_{gas,exp}=0.28\times10^{14}\msol$, one sees that
with $\beta=0.6$ the masses are already
much less exceptional. Also the gas mass fraction with 9\% would be
much closer to the expected values. 
Only the relative gas extent  $E_{\beta=0.6}=0.85$ would still be
unexplainable. Further investigation of this interesting cluster is
definitely necessary. 

In this work we showed that important relations in distant clusters
can be found by a combination of ROSAT and ASCA data. But we also
showed that there are limitations in these data for distant
clusters. We could not find any significant 
evidence for evolution. But on the other hand, due to 
the large uncertainties in the mass determination, in
the temperature and the metallicity measurements we could not rule out
differences between distant and nearby cluster.
We are expecting huge improvements of the situation with the upcoming
X-ray missions XMM, CHANDRA and ASTRO-E. In terms of cosmology with
distant clusters clearly two ways are to be followed. (1) The cluster
parameters used in this work need to be measured with much better
accuracy for clusters with redshifts considered here. In particular, a
more accurate temperature measurement is essential for the
determination of the total mass and the gas mass
fraction. Metallicities with much smaller errors are required to
explain the origin of the intra-cluster medium. 
(2) As the evolutionary effects become stronger with larger redshifts it is
crucial to find clusters at very high redshifts ($z \grsim 1$) like first
studies have shown is possible (e.g. Dickinson 1996; Deltorn et
al. 1997; Carilli et al. 1998).

\begin{acknowledgements}
I am grateful to Doris Neumann for many helpful comments and
for providing PSF-deconvolved fits.
It is a pleasure to thank Chris Collins and Doug Burke for
enlightening discussions and 
Carlo Izzo for his most helpful EXSAS support.
I am also grateful to the ROSAT and the ASCA team for their excellent
support over many years.
\end{acknowledgements}
%
%________________________________________________________________
%

\end{document}